\documentclass[lettersize,journal]{IEEEtran}
\usepackage{amsmath,amsfonts}
\usepackage{algorithmic}
\usepackage{array}
\usepackage[caption=false,font=normalsize,labelfont=sf,textfont=sf]{subfig}
\usepackage{textcomp}
\usepackage{stfloats}
\usepackage{url}
\usepackage{verbatim}
\usepackage{graphicx}
\def\BibTeX{{\rm B\kern-.05em{\sc i\kern-.025em b}\kern-.08em
    T\kern-.1667em\lower.7ex\hbox{E}\kern-.125emX}}
\usepackage{balance}
\usepackage{booktabs}
\usepackage{multirow}
\usepackage{makecell}
\usepackage{cite}
\usepackage[sort,numbers]{natbib}

\begin{document}
\title{An Information-theoretical Secured Byzantine-fault Tolerance Consensus in Quantum Key Distribution Network}
\author{Yi Luo, Hao-Kun Mao, Qiong Li, Member, IEEE
\thanks{Yi Luo, Hao-Kun Mao, Qiong Li was with the
	Department of the School of Computer Science and Technology, Harbin
	Institute of Technology, Harbin, China.
	E-mail: qiongli@hit.edu.cn}}

\markboth{Journal of \LaTeX\ Class Files,~Vol.~18, No.~9, September~2020}%
{}

\maketitle

\begin{abstract}
Quantum key distribution (QKD) networks is expected  to provide information-theoretical secured (ITS) communication over long distances. QKD networks based trusted relay architecture are now the most widely used scheme in practice. However, it is an unrealistic assumption that all relays are fully trustable in complex networks. In the past, only a few studies have theoretically analyzed the case of passive eavesdropping attack by dishonest relays and corresponding defense method. However, we have found that active attacks by dishonest relays can be more threatening. With the consideration of passive and active attacks, we treat dishonest relays as Byzantine nodes and analyzes the upper limit of Byzantine nodes that the QKD network can accommodate. In this paper, we propose an ITS Byzantine-fault tolerance (BFT) QKD network scheme to achieve end-to-end key distribution based on point-to-point QKD links. To ensure consistency and provide BFT ability in the QKD network, we design an ITSBFT-consensus protocol for this network scheme. To ensure the information-theoretic security of consensus, we design a temporary signature scheme based on point-to-point QKD link keys. To prevent Byzantine nodes from disrupting the execution process of key distribution, we design an end-to-end key distribution scheme combined with consensus. We theoretically analyze proposed ITSBFT-QKD network scheme from four aspects: QKD key distribution security, temporary signature security, consensus security, and leader election fairness. The simulation result proved the feasibility and demonstrate the performance.
\end{abstract}


\section{Introduction}
\IEEEPARstart{Q}{uantum}  key distribution is a technique that uses the properties of quantum mechanics to ensure the information-theoretical security\cite{Bennett_Brassard_2014}. In order to further extend the application of QKD, many researches have embarked on the construction of QKD networks. However, the communication distance of point-to-point QKD is limited by the physical devices, and the implementation of long-range end-to-end QKD relies on repeaters (quantum relays\cite{Elkouss_Martinez-Mateo_Ciurana_Martin_2013} or trusted relays \cite{Peev_Pacher_Alleaume_Barreiro_Bouda_Boxleitner_Debuisschert_Diamanti_Dianati_2009} to extend the distance. Given that quantum relays (i.e., devices that can forward quantum bits without measurement or cloning) are still difficult to implement in reality QKD networks, the more practical trusted relaying scheme is widely used adopted to build QKD networks in previous research and several successful demonstrations of QKD networks based on trusted relays have been made\cite{Sharma_Agrawal_Bhatia_Prakash_Mishra_2021}.

In QKD networks based on trusted relays, all relays are assumed to be trusted, which is an unrealistic assumption in complex networks. Once a relay cooperates with an attacker, the secret information of other nodes would be disclosed. Besides, we lack an effective scheme to determine whether relays are attacked or controlled which leads to a serious security threat.

Previous research on weakening the trust assumption focuses on two aspects. On the one hand, physical devices that do not require the assumption of trust can be used to replace trust relay nodes. For example, \citet{Cao_Zhao_Li_Lin_Zhang_Chen_2021} proposed a hybrid trusted/untrusted network architecture based on MDI-QKD, which reduces the number of trusted relays with the advantage that MDI-QKD measurement devices do not need to be trusted. At the same time, they also point out that it is still necessary to introduce trusted relays to extend the QKD distance since it is not allowed to connect two untrusted relays directly \cite{Cao_Zhao_Li_Lin_Zhang_Chen_2021}. On the other hand, a key distribution scheme that can protect key security in the existence of attackers has been proposed. \citet{Desmedt_Wang_2002} discussed that if there are $f$ Byzantine nodes in classical network, theoretically at least $2f+1$ disjoint paths are needed to ensure both security and reliability. \citet{ref1} proposed a multi-path key distribution approach. It pointed out that at least $f+1$ disjoint paths are required for key distribution to guarantee the information-theoretic security of the key. Although the number of necessary paths is reduced, this approach does not guarantee the success rate of distribution. Once an attacker introduces an error, the key distribution process would be aborted .

In practice the attacker's target is not only the key, but also other information. Besides, the attacker can make more active attacks with the keys in their possession. For example, the multi-path scheme cannot defend against following active attacks:
\begin{itemize}
\item{{\bf{Resource contention}: } the attacker makes some key requests to contend for the key resources of other nodes. It makes these resources unavailable for the key distribution of honest nodes, resulting in the number of paths not satisfying the $f+1$ condition.} 

\item{{\bf{Forged requirements: }}the attacker forwards a forged key distribution requirements to other nodes, so that other nodes mistakenly believe that the key distribution is performed with themselves and do not continue to forward the message, making it impossible for the key distribution to proceed properly.}

\item{{\bf{Forged routes: }}Attackers report forged link states to other nodes, send inconsistent routing messages, and interfere with routing protocol work.}
\end{itemize}

Compared to key eavesdropping, these active attacks are more harmful. Thus, we need to not only guard against the attacker's eavesdropping on the key information, but also prevent the attacker from using the key to gain further advantage. 
Compared with previous studies, we make stricter assumptions about the attacker. We assume that there is a part of relay nodes controlled by the attacker Eve. These nodes are refered to as Byzantine nodes which can use owned resources to carry out any attack actively or passively. 

 Our goal is achieve end-to-end key distribution based on multiple point-to-point QKD link keys. However, our task is more difficult due to the introduction of Byzantine nodes. The trust of the QKD system is all protected by point-to-point keys in the original trusted relay network. A node only knows the state of the nodes connected to it, but does not know the distribution of key resources and the link state of other nodes in the network without the forwarding information from other nodes. When all nodes in the network are trusted, the state of non-adjacent nodes can be obtained from the information of other nodes. However, when there exists a certain number of dishonest nodes in the network, the node may get completely different results from the messages of different nodes. To solve this problem, we propose a new mechanism to eliminate the confusion in the QKD network against Byzantine nodes, so that the whole network can agree on the key distribution process and complete the end-to-end key distribution.

Our scheme proposes a new way of establishing and transmitting trust for QKD networks. We use Byzantine fault tolerance\cite{Castro_1999} consensus in QKD networks and  apply point-to-point QKD keys to the consensus to form an information-theoretic secure consensus. This new network scheme enables secure key distribution in QKD networks in the existence of Byzantine nodes. In our scheme, each node in the network needs to agree on a key distribution proposal, which includes the path of key distribution and the number of keys on each path. In addition, since the QKD networks pursue information-theoretic security, this paper proposes a mechanism for constructing information-theoretic secured temporary signatures using QKD keys. Besides, we make improvements on the original multi-path key scheme. we combine the key distribution process with consensus and do our best to ensure the correctness of key distribution. Our main contributions are as follows:
\begin{enumerate}
\item{This paper first presents the  analysis of the betrayed relay nodes as Byzantine nodes with consideration of passive and active attacks. Furthermore, we propose an ITSBFT-QKD network scheme that can accommodate $ MIN\left( C-1,\lfloor \frac{N-1}{3} \rfloor \right)  $ Byzantine nodes, where C is the node connectivity of the network and N is the number of nodes in the network. We proposed an ITSBFT-consensus, which provides ITS in synchronous BFT consensus protocol that helps honest nodes to agree on the key distribution proposal.}

 \item{We propose a temporary signature scheme based on QKD system keys. This scheme adopts QKD keys to provide an information-theoretic secure signature mechanism for the consensus scheme, which helps honest nodes to confirm the source of messages and provide security for the consensus scheme.}
 
 \item{We propose a new end-to-end key distribution scheme with $f+1$ disjoint paths in combination with BFT consensus. Compared to the literature \cite{ref1} does not guarantee delivery, our scheme makes the best effort to guarantee the identity of key distribution.}
\end{enumerate}

\section{The Proposed Framework}
\subsection{ Architecture }
The main purpose of our ITSBFT-QKD network scheme is to solve the problem of end-to-end key distribution with based on point-to-point QKD links against Byzantine nodes. To achieve this purpose, our scheme requires the network to satisfy the following two conditions:
\begin{enumerate}
\item {{\bf{Number of Byzantine nodes:}} Our scheme requires that the number of Byzantine nodes $f$ in the QKD network is less than $ MIN\left( C-1,\lfloor \frac{N-1}{3} \rfloor \right)  $  , where C is the node connectivity of the network and N is the number of nodes in the network. We will analyze this point in our security analysis. }

\item {{\bf{ Synchronization communication:}} if there is no interference from malicious attackers, there is a known upper bound $ \varDelta  $ on the propagation delay of any message in the network, which is same as standard synchrony model assumption in \cite{Abraham2020Sync} }
\end{enumerate}

 In our scheme, messages with propagation delay exceeding $ \varDelta  $ are untrusted, and each node determines the validity of the message according to its own timer. Fortunately, it is not difficult to determine whether a message is time-out because each QKD link in the QKD network requires time synchronization \cite{Wang_Chen_Guo_Yin_Li_Zhou_Guo_Han_2012}\cite{Pljonkin_Rumyantsev_Singh_2017}, which is more accurate compared to classical network time synchronization. In addition, since the connectivity of the network is greater than the number of Byzantine nodes, there are $f+1$ disjoint paths between any two nodes and the nodes can determine the synchronisation status by cross-referencing all paths.

\begin{figure*}[!t]
	\centering
	\includegraphics[width=5in]{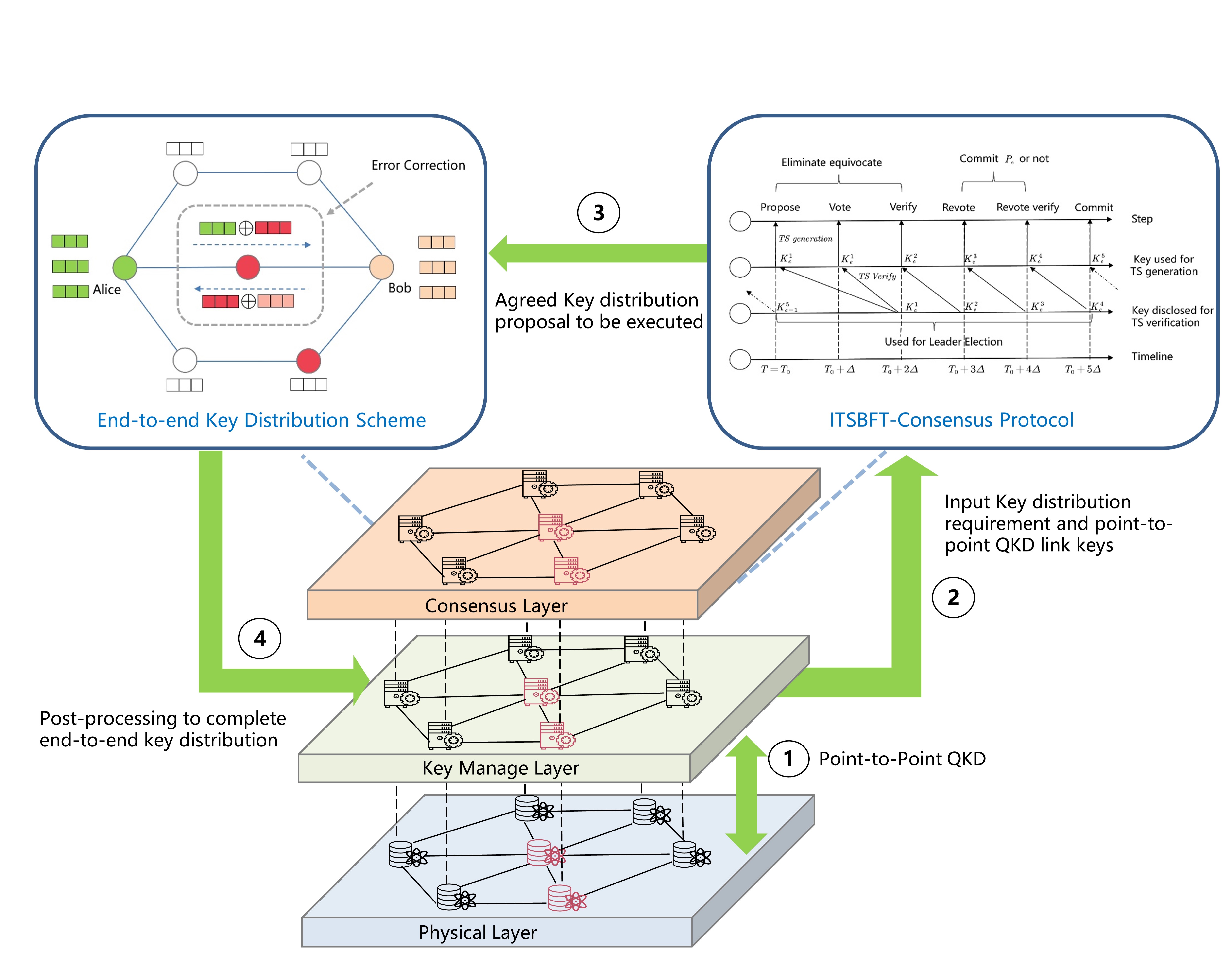}
	\caption{Overview of the proposed ITSBFT-QKD network.}
	\label{fig1}
\end{figure*}

 As shown in the figure \ref{fig1}, there are four steps to complete the key distribution process in our scheme.
 \begin{enumerate}
 \item{ Each adjacent node generates a key via a point-to-point QKD link between them.}
 \item{Each node participates in consensus and agrees on a key distribution proposal.}
 \item{ Each node executes the agreed key distribution proposal.}
 \item{ Perform post-processing on the executed result to generate the final end-to-end key.}
 \end{enumerate}

Steps 1 and 4 belong to the key management layer and can be referred to as existing QKD schemes. Steps 2 and 3 belong to the consensus layer, and we will focus on these steps in the following subsections.
 
 \subsection{ ITSBFT-Consensus workflow } 
\begin{figure}[!t]
  	\centering
  	\includegraphics[width=3.5in]{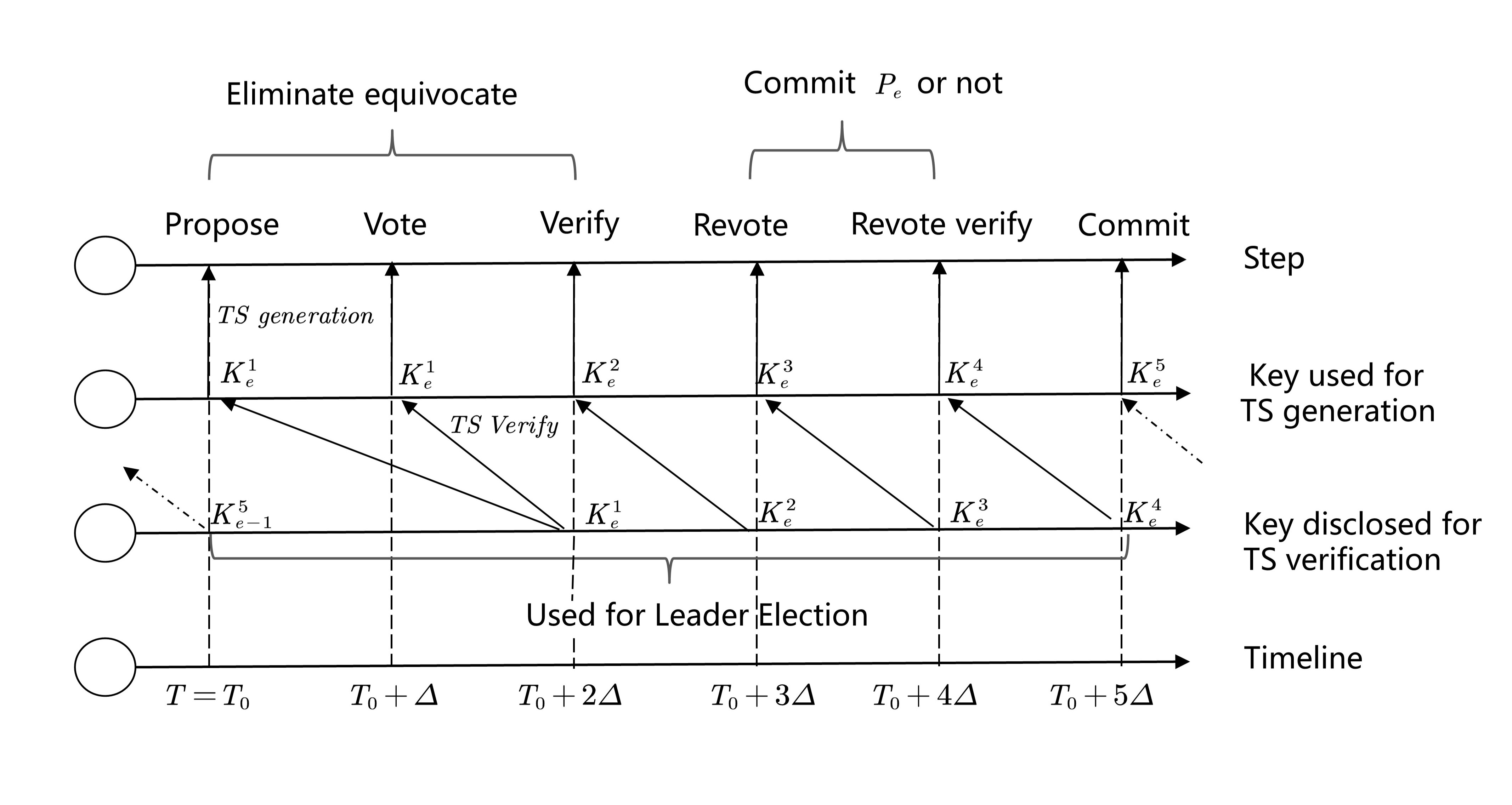}
  	\caption{Workflow and key schedule of ITSBFT-consensus.}
  	\label{fig2}
  \end{figure}

 This section describes the workflow of the proposed ITSBFT-Consensus, which is a synchronous BFT consensus scheme. In our consensus scheme, the leader is responsible for proposing the key distribution proposal, and the leader's term is identified by the view number. To ensure fairness, each view changes its leader based on the TS keys disclosed in the previous view, and a view change is triggered if the leader proposes two equivocate proposals or fails to commit a proposal within the set time. Any node that collects $f+1$ vaild view change messages will move to the next view.
 As shown in the figure \ref{fig2}, a view contains 6 steps, in which all replicas broadcasts a message to participate in the consensus. The temporary signatures of the broadcasted messages are generated using different TS keys, and messages broadcasted at time $T_0$ are verified with the TS keys disclosed at time $T_0+\varDelta  $.

  {\bf{Message Format: }}Messages in the network are broadcast in format $ \left< Step,M,e,TS\left( Key_S \right) ,Key_V \right> _{Source} $. Step represents the Workflow token, M represents the message to be broadcast, e represents the view  number which the message belongs, TS is the Temporary Signature of the message. $ Key_S $ is the key used for the temporary signature, and $ Key_V $ is the key used to verify the last signature.
  
 {\bf{Initial settings: }}Each node keeps the topology and the number of nodes in the network.  While the consensus protocol is in progress, a timer of length $ 8\varDelta  $ is set for each leader. The workflow of ITSBFT-consensus is described as follow:

   \begin{enumerate}
 \item{{\bf{ Propose: }} When a node is selected as the Leader, it needs to form a legitimate proposal and then broadcast $ \left< Propose,P_e,e,TS\left( K_{e}^{1} \right) ,K_{e-1}^{5} \right> _L $. A legitimate proposal is defined as: the number of disjoint key distribution paths is greater than $f$ and the amount of keys distributed on each path is equal.}
 \item{{\bf{Vote: }}When a replica receives the $propose$ message, if the key distribution proposal  $P_e$ is legitimate, vote for the proposal by broadcasting a voting message with the TS of the proposal $ \left< Vote,\left< P_e,TS\left( Proposal \right) \right> ,e,TS\left( K_{e}^{1} \right) ,K_{e-1}^{5} \right> _r $ . }
 \item{{\bf{Verify: }}If the number of Votes received is less than $2f+1$, then a view  change message $ \left< View change,TS\left( K_{e}^{i} \right) ,K_{e}^{i-1} \right> _r  $ is broadcast, otherwise each node will broadcast a verification message $ \left< Verify,TS\left( K_{e}^{2} \right) ,K_{e}^{1} \right> _r $.}
 \item{{\bf{Revote: }}Each node check whether the received TS of the leader is correct and whether there are equivocate leader signatures. If two equivocate TSs are found, then a view  change message is broadcast with the equivocate TSs.$ \left< View change,\left( TS_L,TS'_L \right) TS\left( K_{e}^{i} \right) ,K_{e}^{i-1} \right> _r $. If a node finds that the $ P_e$ it voted for is not correct, the node can change its vote, otherwise it maintains its original vote. Each node broadcast a new vote $ \left< \text{Re}vote,B_e,e,\left( TS_{pre} \right) ,TS\left( K_{e}^{3} \right) ,K_{e}^{2} \right> _r $. }
  \item{{\bf{ Revote verify: }}If at least $2f+1$ votes are received, each node will broadcast $ \left< \text{Re}vote verify,TS\left( K_{e}^{4} \right) ,K_{e}^{3} \right> _r $. }
 \item{ {\bf{Commit: }}If at least $f+1$ nodes vote for a consistent $P_e$, each node will broadcast $ \left< Commit,P_e,e,TS\left( K_{e}^{5} \right) ,K_{e}^{4} \right> _r $. }
\end{enumerate}

 \subsection{Temporary Signature }
 The Temporary Signature scheme consists of three functions: TS key generation, TS generation, and TS verification.It is worth noting that TS has a valid time $\varDelta$ and TS received outside the valid time should not be trusted. 
 
   \begin{figure}[!t]
 	\centering
 	\includegraphics[width=3.5in]{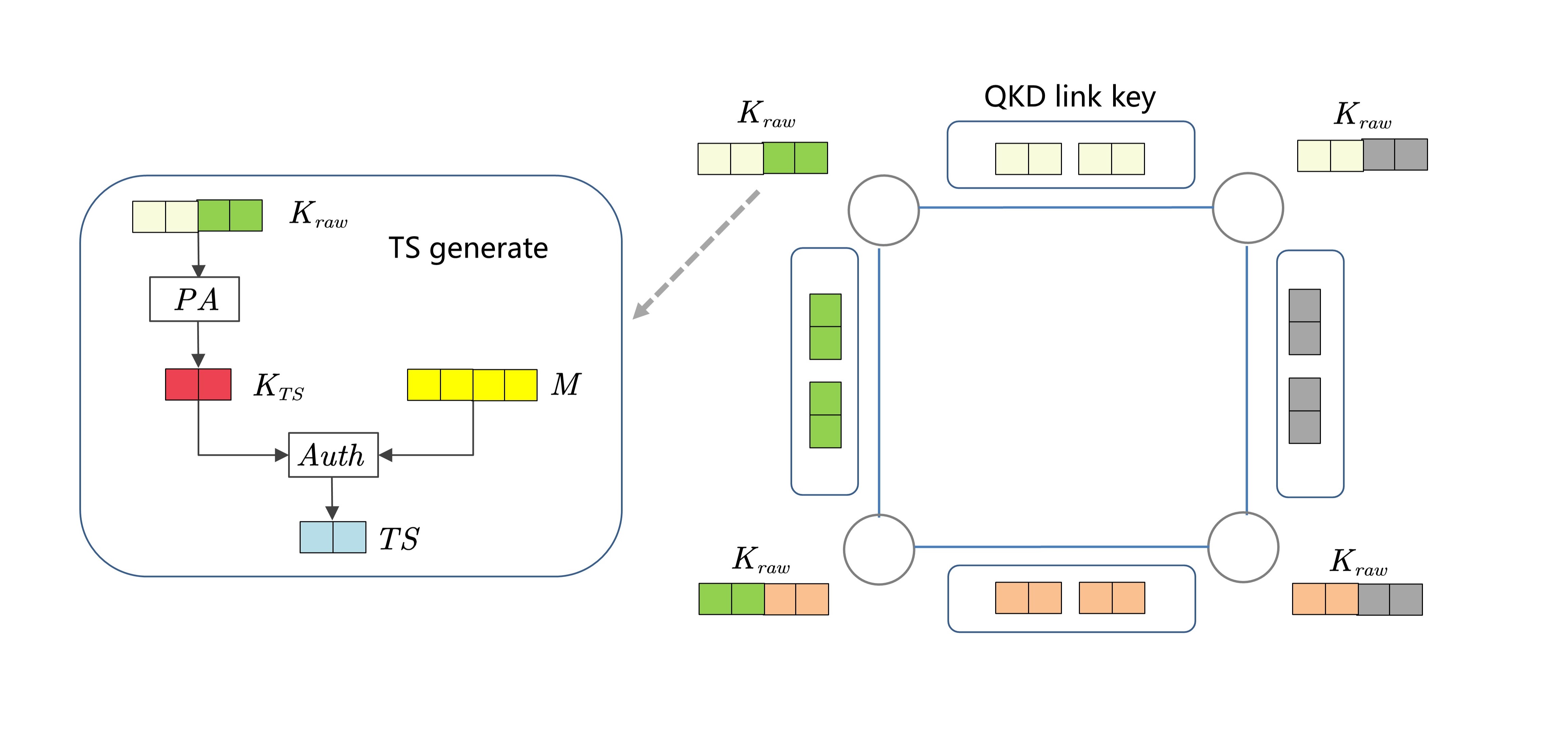}
 	\caption{ Generation process of TS key and TS.}
 	\label{fig3}
 \end{figure}

 {\bf{TS key generation:}} At time $ T_0 $  , each node V obtains $ K_{raw} $  from the neighboring nodes. As shown in figure \ref{fig3}, suppose node V has x neighboring nodes, then the length of the key obtained from each node is $ \frac{L}{x-k} $ , and then it is compressed into a secure key $ K_{TS} $ with length L using privacy amplify. None of the nodes in the network except V can forged $ K_{TS} $ .
 
 {\bf{TS generation:}} At time $ T_0 $, TS can be generated using any of the information-theoretic secure authentication schemes,noted as  $ Auth\left( K_{TS},M \right)$. M is the message to be signed and the result of authentication is temporary signature for $ T_0-T_0+\varDelta  $  time.
 
 {\bf{TS validation:}} At the time $T_0+\varDelta$, all nodes can be validated after $ K_{TS} $ used in time $ T_0 $ diclosed, and the validation needs to satisfy the following two conditions.
 \begin{enumerate}
 \item{ $ Auth\left( K_{TS},M \right) =TS $}
 \item{ There are at least $ f+1 $ disjoint paths between the sender and the receiver of the message in the graph formed by the disclosed key.}
  \end{enumerate}

\subsection{End-to-end Key distribution scheme }
We design a key distribution scheme to assist each node in end-to-end key distribution, which is executed after each node reaches consensus on the key distribution proposal. In our scheme, the key distribution will be done with the help of key Closures.

{\bf{Key Closures:}} Key Closures are xor result of the keys on the input and output links of distribution path.

{\bf{Key Closure transmission:}}
Perform Xor operations on all KCs on a key distribution path except for the source node and destination node.

   \begin{figure}[!t]
 	\centering
 	\includegraphics[width=3.5in]{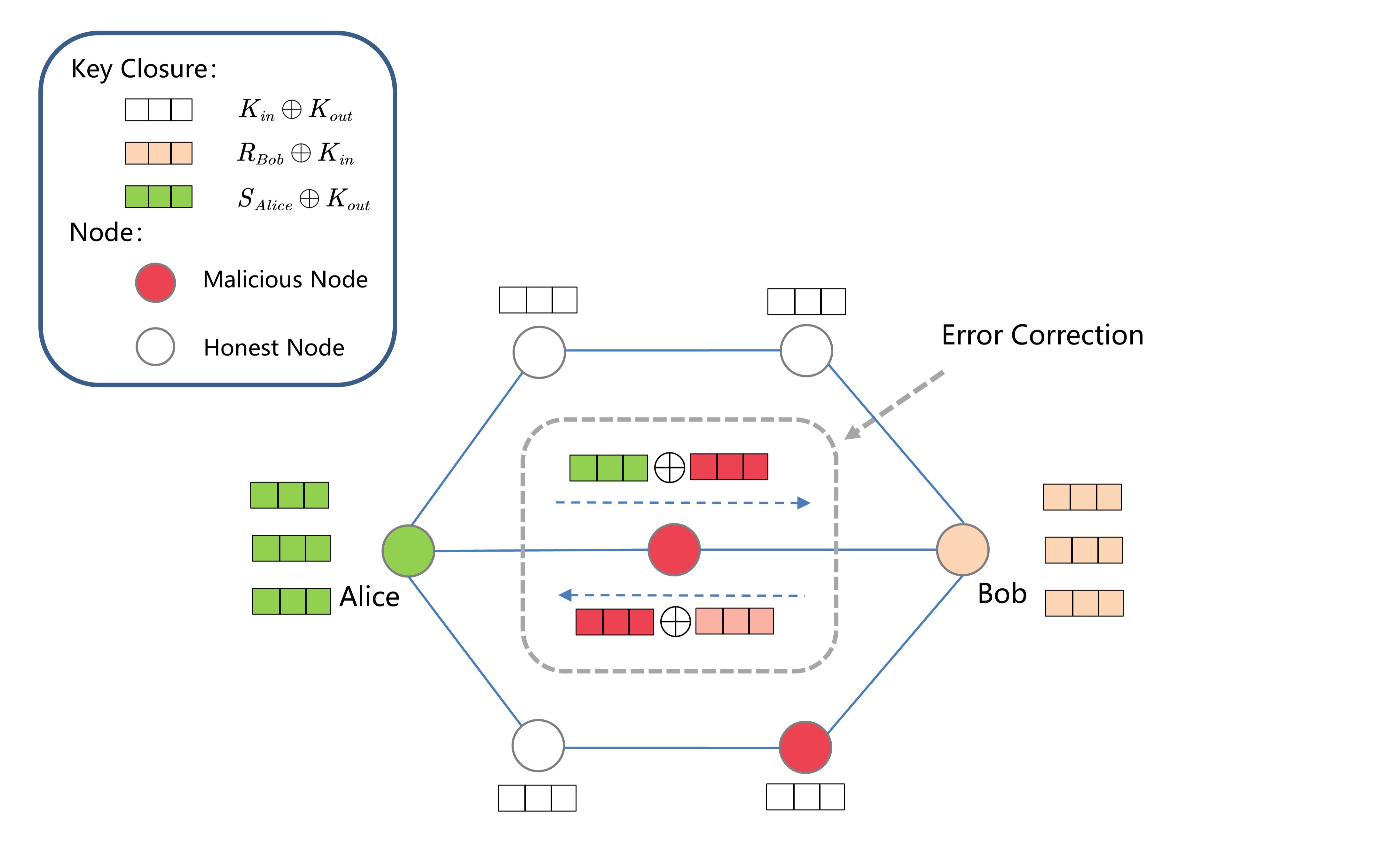}
 	\caption{An instance of proposed end-to-end key distribution scheme.}
 	\label{fig4}
 \end{figure}

End-to-end Key distribution scheme based on ITSBFT workflow. We need to make changes at the $commit$ step and add a round of broadcasts to verify the KCs. we describe the scheme as follow.
\begin{enumerate}
\item{ {\bf{Commit: }}If at least $f+1$ nodes vote for a consistent $P_e$, all nodes except the source and destination nodes generate KC corresponding to its task and broadcast it in $commit$ step  $ \left< Commit,P_e,e,KClsit,TS\left( K_{e}^{5} \right) ,K_{e}^{4} \right> _r $. }
\item{ {\bf{KC verify: }}Each node calculates the calibration results of KC transmission according to received message. then broadcast $ \left< KC verify,P_e,e, Calibration, TS\left( K_{e}^{6} \right) ,K_{e}^{5} \right> _r $.}
\item{If at least $f+1$ consistent calibration results are received and all KC's TS are verified. The source node and the destination node take all KCs as the key distribution result. The source node and the destination node can perform post-processing to complete end-to-end key distribution.}
\end{enumerate}

Our scheme makes maximum efforts to ensure correct key distribution. Under the condition of successful consensus, the honest nodes in the network have reached a consensus on the KC transmission. When the source and destination nodes find that they cannot agree on the key on a path, it is considered that there is a byzantine-fault on this path. Since $f+1$ nodes in the other network have confirmed the KC transmission on the path, the error can be eliminated by distributing the key with the same KC transmission in both directions. The key on this path is considered to be exposed and the attacker can be traced later based on the erroneous KC signature.

\section{Security analysis}
This section focuses on analyzing how our scheme achieves the four properties in the QKD networks. We will analyze information-theoretical security key distribution, ITS temporary signature, BFT consensus security and liveness, leader election fairness. 

{\bf{Information-theoretical security of key distribution: }} Any two nodes A and B in the network perform key distribution, when $ K_A=K_B $, the attacker's mutual information about the distributed keys is less than $ \varepsilon  $.

The security of the key is related to the number of disjoint paths between A and B. the number of attackers  $f$ is less than the number of paths $ \beta   $, in the worst case, there is one attacker on each path, when the ratio of information that the attacker can eavesdrop is $ f/\beta   $. After the privacy amplify process, the mutual information of the attacker obtained can be compressed to less than  $ \varepsilon  $.

It is worth noting that the node connectivity of a QKD network is closely related to the number of disjoint paths. Specifically, if the attacker controls $f$ nodes, then even if these nodes are removed, the network is still connected and key distribution can still be performed. If the number of attackers is greater than $f$, then in the worst case, the set of attacker nodes can form a cut-set of nodes that disconnects A from B, when there is no path between A and B that does not contain attackers.

{\bf{Information-theoretical security of temporary signatures:}} Our scheme guarantees the same security as using the authentication scheme under synchronization conditions when there are $f+1$ disjoint paths between the attacker and the receiver.

  $ K_{TS} $ is completely confidential until it is disclosed, so the probability of an attacker forging the TS is equal to that in the authentication scheme. However, if there are less than $f$ disjoint paths, then it is possible for the attacker to obtain $ K_{TS} $ . e.g., in the worst case where all $f$ neighboring nodes around the receiver are Byzantine nodes, if only $f$ paths are required, then these Byzantine nodes can join together to construct a fake signature.
  
{\bf{Consensus safety and liveness:}} No two honest nodes will commit equivocate key distribution proposals. In the case that the topology's node connectivity satisfies the condition, the honest node during the consensus process will not form a deadlock and will definitely output $ P_e $   or  $ \bot  $.

{\bf{Safety: }}In our scheme honest nodes will only vote for the legitimate key distribution proposal. Duplicate TSs will be found after the key is disclosed. In the first vote step, each node broadcasts the proposal it voted for. So in the revote step, all honest nodes can determine if there are two equivocate proposals. Since there are only $f$ attackers, at least one honest node needs to vote for the key distribution proposal to get $f+1$ votes on consistent proposal, so eventually there must be agreement on a legitimate proposal or $ \bot  $. 

{\bf{Liveness: }}It is important to note that since an attacker can block network communication or cause a QKD interruption through eavesdropping.  Our consensus scheme cannot guarantee a continuous output of valid proposals. When the QKD network does not satisfy the conditions for reaching consensus, the consensus output at that point is $ \bot  $.
In general,the leader needs to commit a valid proposal within $8\varDelta $. If the malicious leader makes an illegal proposal or broadcast equivocate proposals, then $ \bot  $ is output and move to the next view.  

{\bf{Fairness:}} Any node has equal probability of becoming a leader. 

In the process of leader election, each leader has equal probability of being selected and each leader consumes a limited amount of key resources. The next leader is generated by the QKD key used in the previous round. Due to the quantum randomness of the QKD key, any node has equal probability of becoming a leader.

\section{simulation}
\subsection{simulation setup}
Our simulations are performed in 3 topologies based on QKDnetsim\cite{Mehic_Maurhart_Rass_Voznak_2017}, as shown in figure 5. In each topologies,  3 experiments are conducted by the number of Byzantine nodes of 1, 2, 3 respectively. Byzantine nodes are located on the disjoint paths to get the maximum advantage.

The QKD link settings are same as the settings in \cite{Mehic2020Novel}. We pre-store 10Mb QKD keys on each link, and the maximum bandwidth of the classical communication channel is 20Mbps.

For the consensus parameter, we set $ \varDelta  $ to 1s, the PA scheme we used in temporary signature is Modified Toeplitz Hash \cite{Hayashi_2011} and the authentication scheme we used is \cite{Kiktenko_Malyshev_Gavreev_Bozhedarov_Pozhar_Anufriev_Fedorov_2020} because its key consumption is more stable at different data lengths, where the parameters are $ \omega =63,\varepsilon _k=10^{-12} $ . The upper limit of key distribution between any two nodes within a single round of consensus is 300Kb. In post-processing we only consider compress the amount of information that the attacker has eavesdropped on.
\subsection{simulation results}
We conducted simulation based on the above settings. The nodes with the same color in the topology have a 500Kb key distribution requirement between them, and The simulation results are shown in table \ref{table1}.
\begin{figure*}[!t]
	\centering
	\subfloat[]{\includegraphics[width=2in]{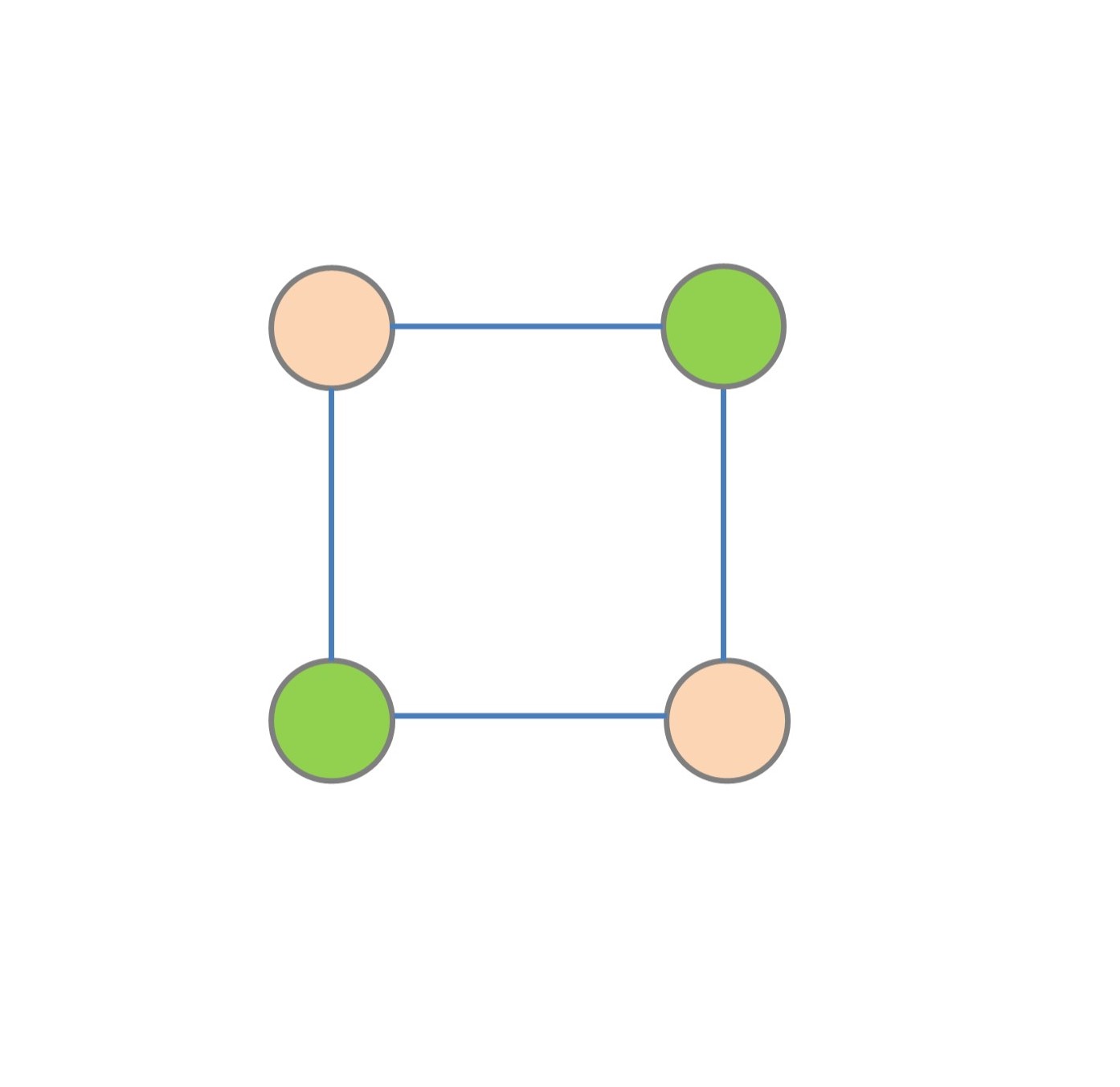}%
		\label{fig_first_case}}
	\hfil
	\subfloat[]{\includegraphics[width=2in]{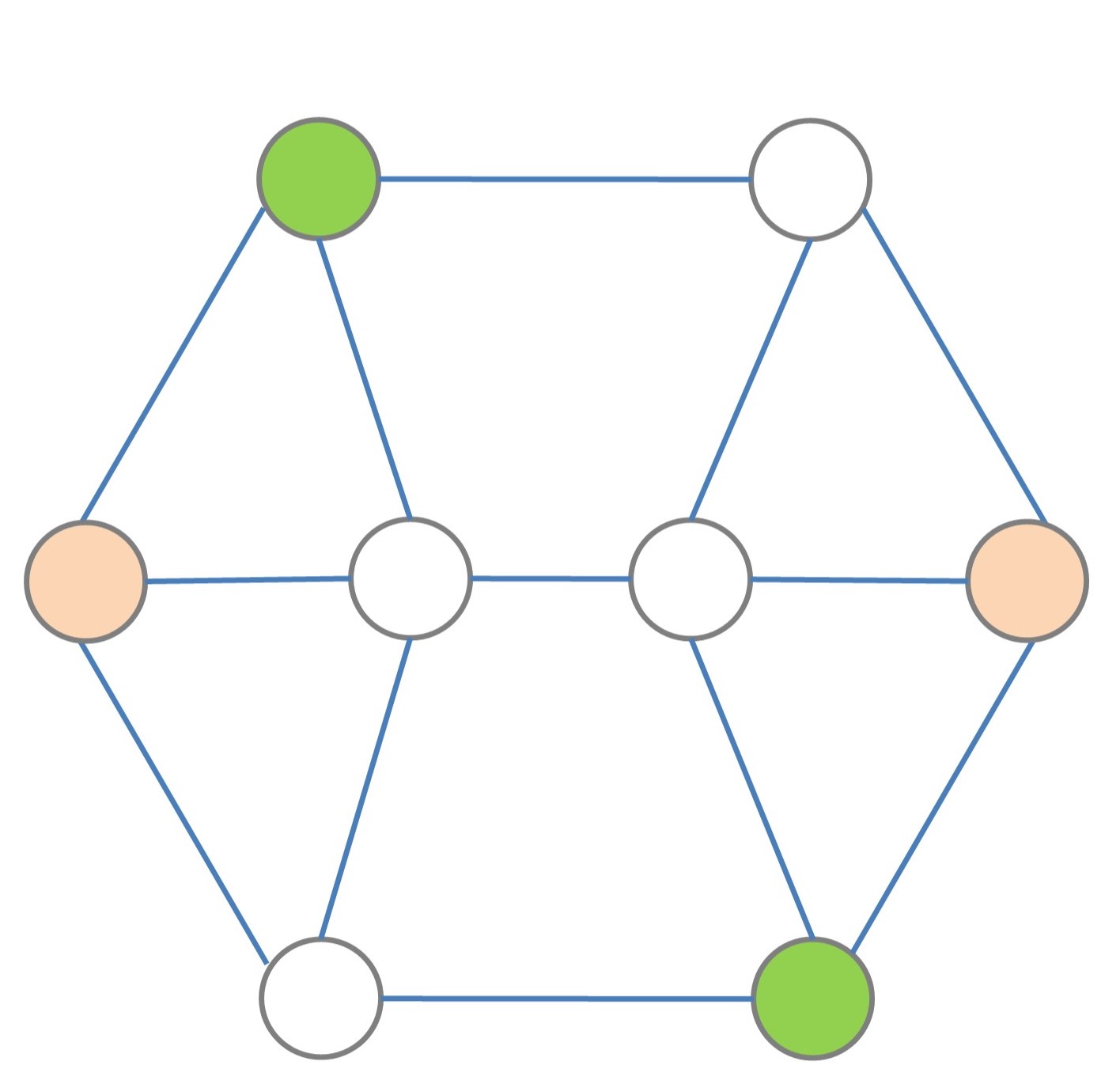}%
		\label{fig_second_case}}
	\hfil
	\subfloat[]{\includegraphics[width=2in]{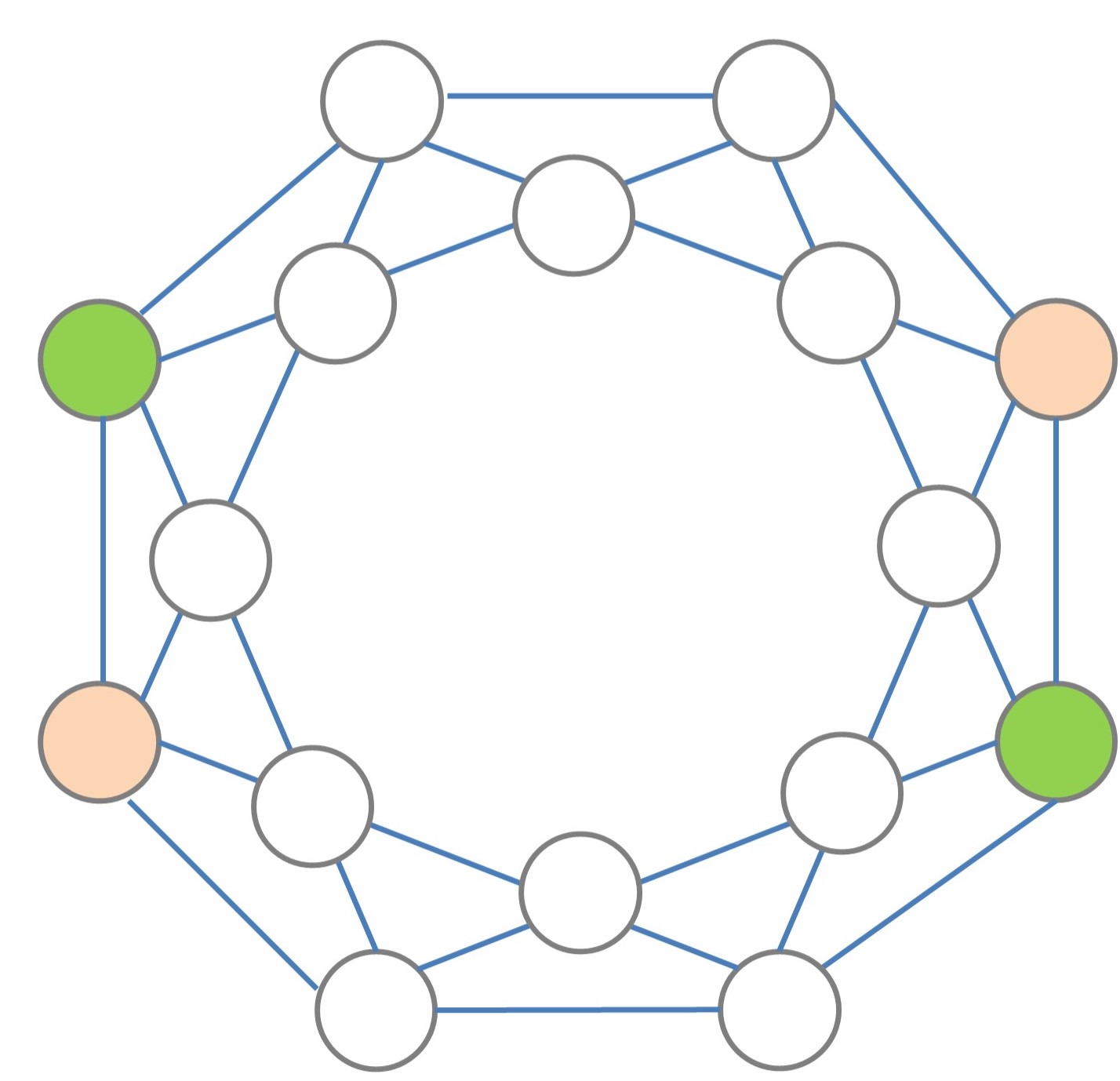}%
		\label{fig_third_case}}
	\caption{Three topologies with different node connectivity are used for the simulation. According to our analysis, the number of Byzantine nodes that can be accommodated in the three topologies is 1,2,3. In the simulation, there is a 500Kb key distribution requirement between nodes of the same color.}
	\label{fig_sim}
\end{figure*}

\begin{table*}[htbp]
	\centering
	\caption{Simulation results in three topologies}
	\begin{tabular}{ccccccc}
		\toprule
		Topology  & \makecell[c]{Number of\\Byzantine nodes} & 
		\makecell[c]{Key consumption\\total network(Kb)} & 
		\makecell[c]{Key consumption\\In consensus(Kb)} & 
		\makecell[c]{Key consumption\\In delivery(Kb)} & 
		\makecell[c]{Percentage of Byzantine\\node eavesdropping}  &
		\makecell[c]{Time(s)} \\
		\midrule
		\multicolumn{1}{c}{\multirow{3}[2]{*}{Figure5-a}} & 1     & 4011.27  & 11.27  & 4000  & 50.00\% & 14 \\
		& 2     & -        & -      & -   & 100.00\% & - \\
		& 3     & -        & -      & -   & 100.00\% & - \\
		\midrule
		\multicolumn{1}{c}{\multirow{3}[2]{*}{Figure5-b}} & 1     & 4516.90  & 16.90  & 4500  & 33.33\% & 7 \\
		& 2     & 9067.59  & 67.59  & 9000  & 66.67\% & 14 \\
		& 3     & -        & -      & -   & 100.00\% & - \\
		\midrule
		\multicolumn{1}{c}{\multirow{3}[2]{*}{Figure5-c}} & 1     & 6030.04  & 30.04  & 6000  & 25.00\% & 7 \\
		& 2     & 9045.06  & 45.06  & 9000  & 50.00\% & 7 \\
		& 3     & 18180.25  & 180.25  & 18000 & 75.00\% & 14 \\
		\bottomrule
	\end{tabular}%
	\label{table1}%
\end{table*}%

According to the simulation results in table \ref{table1}. We list the point-to-point QKD link keys consumption of total network, consensus process and key distribution process. The Time listed in table is represents the delay introduced in the consensus. The percentage of information that Byzantine nodes can eavesdrop shows the ability of different topologies to defend Byzantine nodes. As the node connectivity increases, its capacity to accommodate Byzantine nodes also increases. For example, figure \ref{fig_first_case} with a node connectivity of 2 can only accommodate  one Byzantine node, and when there are more than one Byzantine nodes, both temporary signature and key distribution are no longer secure. In contrast, figure \ref{fig_third_case} with a node connectivity of 4 can accommodate three Byzantine nodes.

Although our proposed ITSBFT-consensus scheme consumes keys continuously, it does not massively consume keys in the QKD network. In our simulation, the keys consumed by the consensus scheme increase with the number of Byzantines set up. In the worst case, such as Figure \ref{fig_third_case} with 3 Byzantine nodes, the key consumed in consensus in this case is at most 1.001\% of the key consumed in end-to-end key distribution. It can be seen that our scheme is feasible since it does not consume many keys. 

\section{Conclusion}
In this paper, we propose an ITSBFT-QKD network scheme that enables end-to-end key distribution against $ MIN\left( C-1,\lfloor \frac{N-1}{3} \rfloor \right)  $ Byzantine nodes. First, we propose a synchronous ITSBFT-consensus process. Second, we proposed the temporary signature to guarantee information-theoretical security of the consensus scheme. Third, based on the consensus of the key distribution proposal, we propose a key distribution scheme to address the errors that may be maliciously introduced by Byzantine nodes. We analyze the advantages of our network architecture in the security analysis, which ensures not only the ITS and fairness of the key distribution scheme but also the safety and liveness of the consensus scheme with the help of QKD keys. Our simulations have proved the feasibility and demonstrate the performance of our scheme.

Besides, we found that BFT consensus could effectively improve the consistency and the fault tolerance ability of QKD networks, and QKD could guarantee higher security for consensus. Therefore, the proposed scheme could be applied to other applications in the future and is expected to become the next generation of security infrastructure.

\bibliographystyle{unsrtnat}

\bibliography{bibfile.bib}

\end{document}